\documentclass[12pt,preprint]{aastex}
\usepackage{emulateapj5,apjfonts}
\input psfig.tex

\newcommand{\etal}{{et~al.}}

\newcommand{\Msun}{M_\odot}

\newcommand{\kms}{$\rm {km}~\rm s^{-1}$}

\begin{document}

\lefthead{NGC~1399}
\righthead{Gebhardt~\etal}

\title{The Black Hole Mass and Extreme Orbital Structure in NGC~1399}

\author{Karl Gebhardt\altaffilmark{1}, 
Tod R. Lauer\altaffilmark{2}, 
Jason Pinkney\altaffilmark{3}, 
Ralf Bender\altaffilmark{4}, 
Douglas Richstone\altaffilmark{5}, 
Monique Aller\altaffilmark{5}, 
Gary Bower\altaffilmark{6}, 
Alan Dressler\altaffilmark{7}, 
S.M.~Faber\altaffilmark{8}, 
Alexei V. Filippenko\altaffilmark{9}, 
Richard Green\altaffilmark{2}, 
Luis C. Ho\altaffilmark{7}, 
John Kormendy\altaffilmark{1}, 
Christos Siopis\altaffilmark{5}, and 
Scott Tremaine\altaffilmark{10}}

\altaffiltext{1}{Department of Astronomy, University of Texas, Austin,
Texas 78712; gebhardt@astro.as.utexas.edu, kormendy@astro.as.utexas.edu}

\altaffiltext{2}{National Optical Astronomy Observatories, P. O. Box
26732, Tucson, AZ 85726; lauer@noao.edu, green@noao.edu}

\altaffiltext{3}{Ohio Northern University; j-pinkney@onu.edu}

\altaffiltext{4}{Universit\"ats-Sternwarte, Scheinerstrasse 1,
M\"unchen 81679, Germany; bender@usm.uni-muenchen.de}

\altaffiltext{5}{Dept. of Astronomy, Dennison Bldg., Univ. of
Michigan, Ann Arbor 48109; dor@astro.lsa.umich.edu,
jpinkney@astro.lsa.umich.edu}

\altaffiltext{6}{Computer Sciences Corporation, Space Telescope
Science Institute, 3700 San Martin Drive, Baltimore, MD 21218;
bower@stsci.edu}

\altaffiltext{7}{The Observatories of the Carnegie Institution of
Washington, 813 Santa Barbara St., Pasadena, CA 91101;
dressler@ociw.edu, lho@ociw.edu}

\altaffiltext{8}{UCO/Lick Observatories, University of California,
Santa Cruz, CA 95064; faber@ucolick.org}

\altaffiltext{9}{Department of Astronomy, University of California,
Berkeley, CA 94720-3411; alex@astro.berkeley.edu}


\altaffiltext{10}{Princeton University Observatory, Peyton Hall,
Princeton, NJ 08544; tremaine@astro.princeton.edu}

\begin{abstract}

The largest galaxies, and in particular central galaxies in clusters,
offer unique insight into understanding the mechanism for the growth
of nuclear black holes. We present {\it Hubble Space Telescope} kinematics
for NGC~1399, the central galaxy in Fornax. We find the best-fit model
contains a black hole of $(5.1 \pm 0.7)\times 10^8~\Msun$ (at a distance
of 21.1 Mpc), a factor of over 2 below the correlation of black hole mass
and velocity dispersion. We also find a dramatic signature for central
tangential anisotropy. The velocity profiles on adjacent sides
0.5\arcsec\ away from the nucleus show strong bimodality, and the
central spectrum shows a large drop in the dispersion.  Both of these
observations point to an orbital distribution that is tangentially
biased. The best-fit orbital model suggests a ratio of the tangential
to radial internal velocity dispersions of three. This ratio is the
largest seen in any galaxy to date and will provide an important
measure for the mode by which the central black hole has grown.

\end{abstract}


\section{Introduction}

It is clear that the mass of the central black hole is related to its
host galaxy in a fundamental way. Dressler (1989), Kormendy (1993),
Kormendy \& Richstone (1995), and Magorrian (1998) were the first to
highlight a correlation between the black hole mass and the bulge
light. Subsequently, many other correlations have been found, with the
tightest being that between black hole mass and velocity dispersion
(Gebhardt et al. 2000, Ferrarese \& Merritt 2000). Numerous
theoretical models have been proposed to explain these correlations,
and the most compelling to date are those that work through active
galactic nucleus (AGN) feedback mechanisms (Silk \& Rees 1998; Fabian
1999; Springel et al. 2005; Robertson et al. 2006). To push further
requires more secure observations. The problem is that the
uncertainties in the black hole mass estimates are still large (around
30-50\%) and, more importantly, the extremes of the correlations are
not well explored. We have been targetting the largest galaxies in
order to study their central black hole mass. There is only a handful
of objects studied with velocity dispersions above 300~\kms.

In this paper we study the giant elliptical NGC~1399, the dominant
galaxy in the Fornax cluster. In addition to providing information on
the upper end of black hole correlations, central galaxies in clusters
offer unique insight. These galaxies are subject to significant
accretion and mergers, and it is important to understand whether the
black hole grows as the galaxy grows. Houghton et al. (2006) study
NGC~1399 using adaptive optics observations on the VLT; they find a
black hole mass of $(1.2 \pm 0.6) \times 10^9~\Msun$ (for a distance
of 19.9 Mpc). We find a black hole mass of $(5.1 \pm 0.7) \times
10^8~\Msun$ (for a distance of 21.1 Mpc as used in Lauer et
al. 2005). While over a factor of two different, we are consistent
within $1\sigma$ (the Houghton et al. result is only a $2\sigma$
significance for a black hole detection). Furthermore, we find
similar results in terms of the central orbital structure. We use a
distance of 21.1 Mpc to NGC~1399 from Tonry et al. (2001), but scaled
to H$_0$=70 as in Lauer et al. 2005.

\section{Data}

\subsection{HST Observations}

The surface brightness profile comes from WFPC2 observations for {\it
Hubble Space Telescope (HST)} programs GO-5990 and GO-8214 (PI:
Grillmair). NGC~1399 was observed for 4000~s in F606W and for 5200~s
in F450W, with the galaxy centered on the PC. The surface brightness,
ellipticity, and color gradient profiles are shown in Fig 1. Due to
NGC~1399 being nearly round, the position angle is very uncertain and
we do not include discussion of it. The reductions are discussed by
Lauer et al. (2005). From its surface brightness profile, NGC~1399 is
classified as a core galaxy with a break radius 3.2\arcsec\ and
$\gamma=0.12$ (approximately the central projected density slope),
from a Nuker Law fit (Lauer et al. 2005). For the surface brightness
beyond the {\it HST} image, we use ground-based imaging from Saglia et
al. (2000). We match the ground-based $R$-band data to the {\it HST}
surface brightness in the overlap region.

Figure 1 shows a variation in the ellipticity inside of 0.3\arcsec.
However, since the surface brightness is not steep in the central
regions and the isophotes are nearly round, there are large
uncertainties in the ellipticities. Thus, the variation could be due
to noise and a constant ellipticity model provides nearly identical
residuals.  In the models that follow, we use a constant ellipticity
of 0.1, but the results do not change much when using an ellipticity
of zero. This is also consistent with the surface brightness at larger
radii. The position angle is $110^\circ$ (measured N to E), and we
assume it to be constant.  With both a constant PA and ellipticity, we
deproject NGC~1399 as in Gebhardt et al. (1996). This deprojection is
used in the dynamical models. From the bottom panel in Fig. 1, there
is essentially no color gradient in NGC~1399. Thus, we use a constant
mass-to-light ratio for the stellar potential. In addition, the color
map is also constant as a function of position angle.

We have checked whether the isophotal centers change as a function of
radius. This check is important for the discussion in Section 4. We
find that the center from isophotes at 10\arcsec\ compared to that
derived from the isophotes in the central regions is consistent to
within 0.2 pixels, or better than 0.01\arcsec. Thus, there appears to
be no deviation in the galaxy center. In addition, we find no evidence
in the residual map (Lauer et al. 2005) for any second
component. Houghton et al. (2006) see an elongation in the central
0.5\arcsec, suggesting a possible eccentric disk. We find no such
structure in our images; furthermore, the {\it HST} images have been
subsampled and deconvolved, giving a FWHM around 0.05\arcsec, better
than the 0.078\arcsec\ as reported for the adaptive optics K-band
image of Houghton et al. (2006). Still, it is difficult to reconcile
the differences; we attribute them to either different structure in
$K$ band versus the $R$ band (however, this is unlikely) or an
adaptive optics artifact. Higher signal-to-noise ratio (S/N) and
repeat adaptive optics observations will likely help determine the
cause.

The central surface brightness of NGC~1399 is $V = 16.0$ mag
arcsec$^{-2}$, making it one of the faintest targets we have observed
with {\it HST}/STIS (Pinkney et al. 2003). For our previous
observations, however, we have used a high-resolution grating centered
on the Ca~II triplet region at 8500~\AA. This region is the best to
use since it is not greatly influenced by stellar template mismatch
and continuum estimation, as are the bluer regions (Barth et
al. 2002). However, the exposure times become prohibitive for targets
fainter than $V = 16.0$ mag, since it typically requires exposure
times of longer than 17 hours to obtain adequate signal. Complicating
the kinematic estimate is that these large galaxies tend to have large
black hole masses, and therefore large central dispersions. The
dispersion of the central STIS pixel for NGC~4649 is over 600~\kms;
given the relatively small equivalent widths of the Ca~II triplet
lines, the large dispersion makes the lines almost disappear into the
continuum. Our strategy for NGC~1399 is to use a lower resolution
grating over the Ca~II H\&K region, where the lines remain clear even
when the dispersion is that high (Dressler 1984).

\vskip 10pt \psfig{file=f1.ps,width=8.5cm,angle=0}
\figcaption[gebhardt.fig1.ps]
{{\it HST} photometry of NGC~1399. The
top panel is the F606W (close to $R$ band) surface brightness profile.
We only show the radial region included in {\it HST} images. The
middle panel is the ellipticity profile. The bottom panel is the
difference in surface brightness between F450W (close to the $B$ band)
and F606W. The spatial resolution is about 0.05\arcsec, so the two
central points are within the resolution element; given the shallow
gradient of NGC~1399, we expect the central two points to not be
biased. There are large ellipticity changes inside of 1\arcsec, which is
a result of the shallow gradient of the surface brightness and relatively
low ellipticity.
\label{fig1}}
\vskip 10pt


We obtained 6.67 hours of observations on STIS (Woodgate et al. 1998)
using the G430L grating with $52\times0.2$\arcsec\ slit align at
$117^\circ$ (along the major axis, see Figure 3 of Lauer et
al. 2005). The wavelength range is 2880--5690~\AA, with 2.746~\AA\ per
pixel. We binned on chip by two, providing 0.1\arcsec\ per pixel in
the spatial direction. This setup gave us S/N $\approx 20$ per pixel
in the central regions, and the same S/N at a radius of 1.6\arcsec\ by
binning over 10 pixels (1.0\arcsec).

With the low spectral resolution and the wide slit, we have to pay
special attention to the change in the instrumental resolution when
observing a point source compared to observing a diffuse source.  We
observed three different template stars with this setup: 
HD141680 (a G8III star), HD165760 (G8III), and HD188056 (K3III). We 
stepped each star perpendicularly across the slit to monitor
the change in velocity centroid. The goal was to create a template
star that represents the actual surface brightness of the galaxy
across the slit. Figure 2 shows the shift in the velocity centroid as
a function of position in the 0.2\arcsec\ and the 0.1\arcsec\ slit.
The peak to peak variation is about 700~\kms\ for the 0.2\arcsec\
slit, which is expected given the 2.746~\AA\ per 0.05\arcsec\ pixel.
Figure 3 plots the relative intensity variation across the slit. This
intensity variation must be taken into account as well when creating a
proper template; at the edges of either slit, there is almost a 50\%
drop in intensity compared to the center.

As a first step, we need to know the actual spectral resolution for
our setup and galaxy. Fortunately, for two of the stars, HD141680 and
HD165760, high-resolution ground-based spectra exist over
our spectral range (Leitherer et al. 1996). Therefore, we can 
compare the high-resolution spectra with 
our spectra to obtain the instrumental
resolution. Since our wavelength region of interest is
3900--4500~\AA, we concentrate on this region only. We do this in three
different ways to demonstrate the extremes of the results. First,
before summing the stepped template, we remove the velocity shift
across the slit which corresponds to a point source. In this case we
find that the instrumental $\sigma_i= 207$~\kms, which is roughly what we
would expect given this setup. Second, we sum the light for the
templates without removing the velocity shift. This case corresponds
to a flat source across the slit, and here we find an instrumental
$\sigma_i = 272$~\kms. In the third case, we include the surface
brightness profile for NGC~1399. There, the light at the edges drops
by 20\%, and we find an instrumental $\sigma_i = 275$~\kms, consistent within
the uncertainties with the flat profile. We can also use the lamp
lines to get an estimate of the instrumental dispersion. For lines in
this wavelength region, we measure an instrumental $\sigma_i =
298$~\kms. However, in the regions of interest, finding isolated lines
is difficult, and $\sigma$ is somewhat overestimated; also, the
lamp lines are a completely flat source unlike NGC~1399. Thus, we use
275~\kms\ as our instrumental $\sigma_i$ (implying a FHWM of 646~\kms).
Given the instrumental resolution of 275~\kms\ for this setup, it is
difficult to obtain an accurate estimate of the galaxy dispersion if
it is below this.

The spectra cover the range 2880--5690~\AA; however, for the kinematics we
use only the region 3850--4400~\AA, covering the Ca~II H\&K lines and
the G-band at 4300~\AA. Below 3850~\AA, the lines are weak and the
continuum drops, making the S/N too low to be
useful. The Mg region around 5100~\AA\ still has good signal, but
there are issues with template mismatch that are difficult to
overcome. Since NGC~1399 has one of the largest dispersions, it will
also have one of the largest equivalent widths for Mg, making it
difficult to find templates that accurately reflect the galaxy.  This
is a longstanding problem, and the traditional method for handling
this is to either fit the kinematics in Fourier space (which removes
the equivalent width difference), or dilute the galaxy equivalent
width by adding a constant to the continuum. Unfortunately, both of
these seek to simply match the equivalent width, and any shape
difference between the galaxy and the template may manifest itself by
biasing the kinematics. We therefore choose to exclude the Mg region
during the fits. Barth et al. (2002) find a similar result when
comparing kinematic result from different spectral regions.


\psfig{file=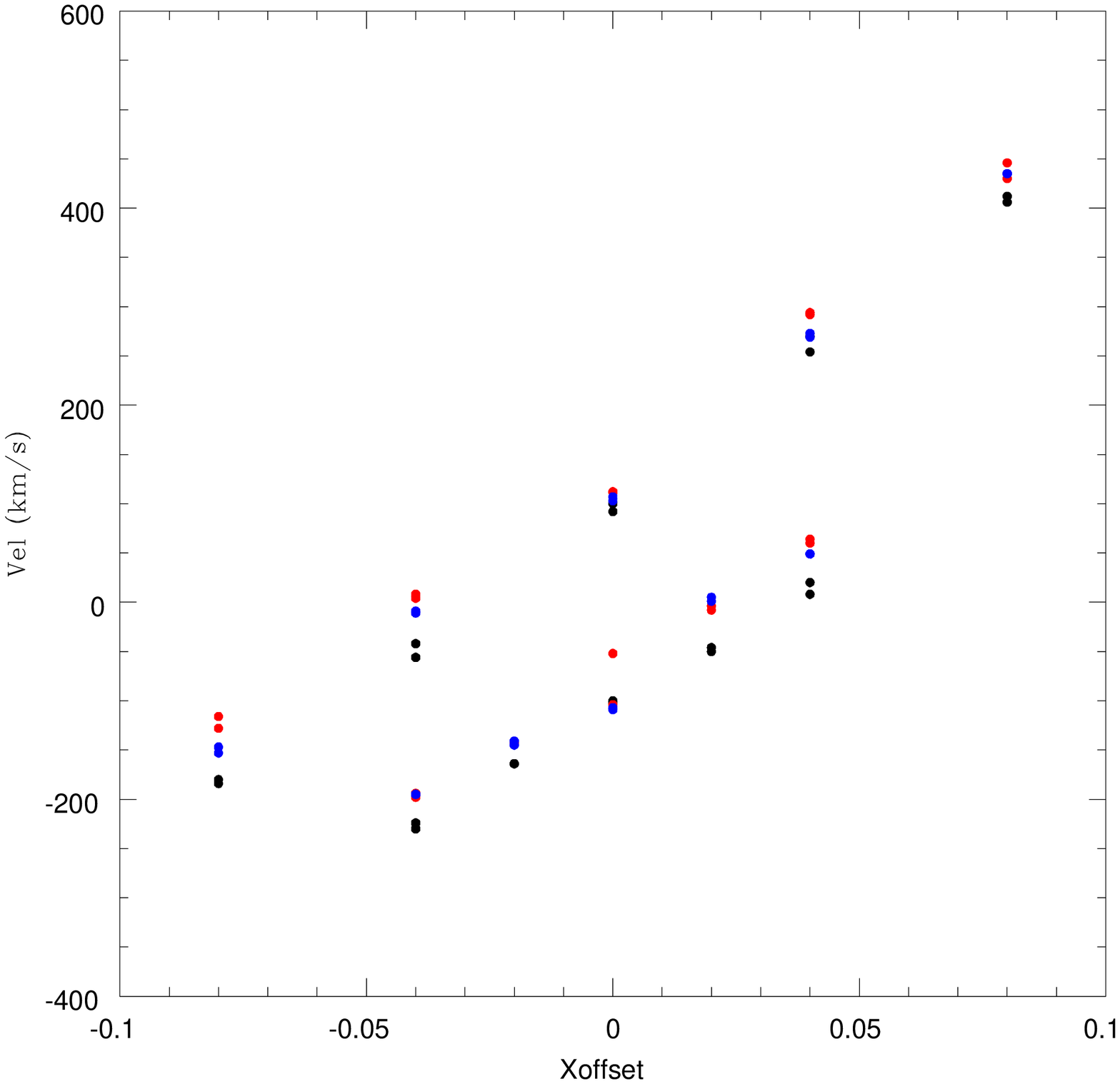,width=8.5cm,angle=0}
\figcaption[gebhardt.fig2.ps]
{The velocity offset as a function of
position across the 0.2\arcsec\ slit (top set of points) and across
the 0.1\arcsec\ slit (bottom set of points). The overall velocity
offset is arbitrary and has been set so that the two sets of points do
not overlap. Each color corresponds to a different star, and each star
was stepped two times across the slit.  The slight velocity
differences at a given spatial position are a function of both the
accuracy in centroiding the velocity and the accuracy in pointing the
telescope. The total velocity shift from end to end is as expected,
given the 2.746~\AA\ per pixel and 0.05\arcsec\ pixels.
\label{fig2}}



\psfig{file=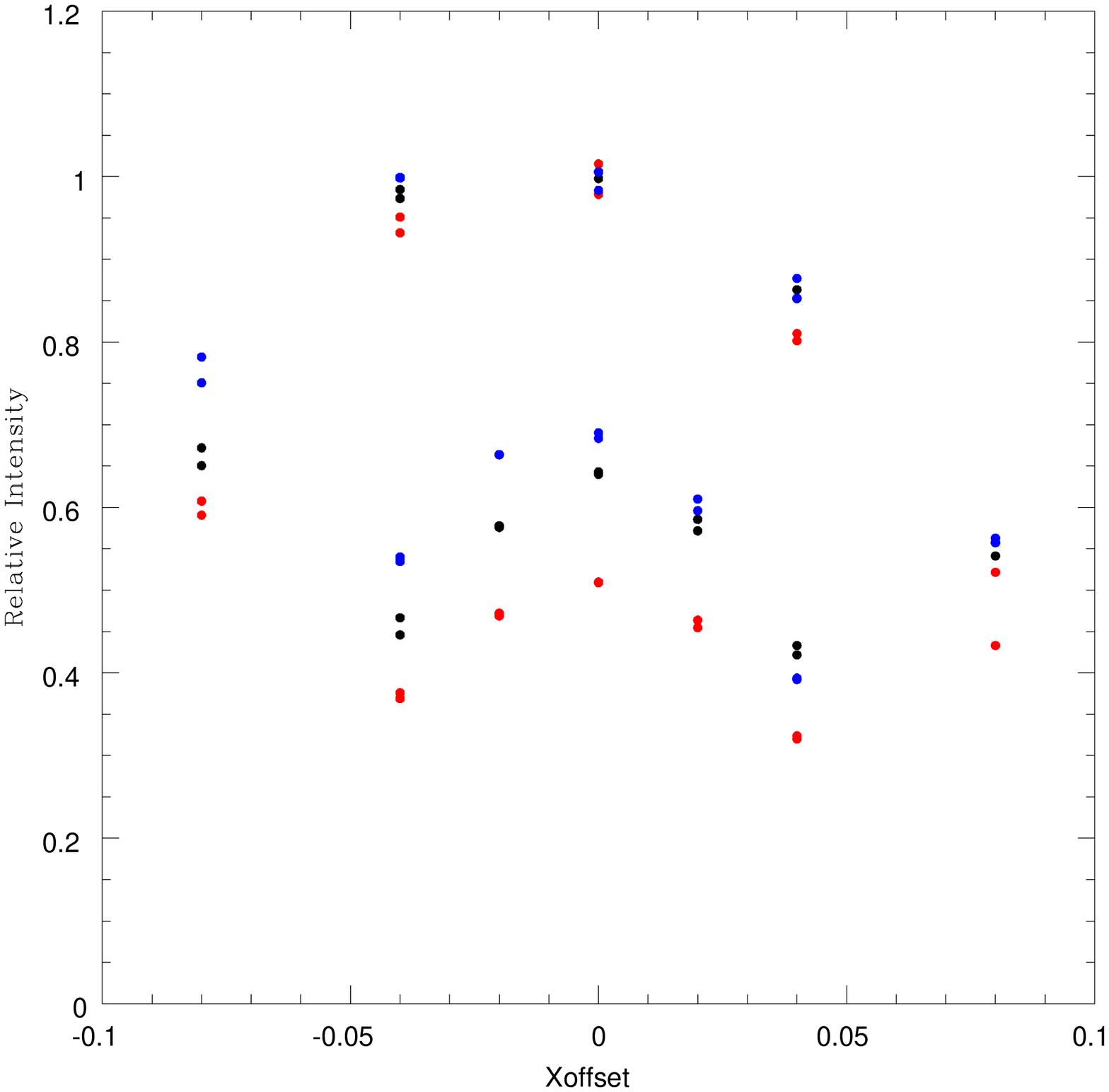,width=8.5cm,angle=0}
\figcaption[gebhardt.fig3.ps]
{The relative intensity as a function of
position across the 0.2\arcsec\ slit (top set of points) and across
the 0.1\arcsec\ slit (bottom set of points). Each color corresponds
to the same star as in Figure 1.
\label{fig3}}
\vskip 10pt


Figure 4 plots the spectra of the central pixel and at a radius of
0.8\arcsec\ for NGC~1399.  The two obvious features are the Ca~II H\&K lines
at 3900~\AA\ and the G-band at 4300~\AA. The template that we use is a
result of the procedure described above. For each of the three
template stars, we sum the light as it was stepped across the slit
with a weight that corresponds to the light profile for NGC~1399. The
fit to the galaxy spectrum then involves a convolution with a velocity
profile and a linear combination of the three templates. We use the
fit as described by Gebhardt et al. (2000) and Pinkney et al. (2003),
where we estimate a non-parametric line-of-sight velocity profile. The
red lines in Figure 4 represents the best-fit velocity profiles
convolved with the template.

\vskip 10pt
\psfig{file=f4.ps,width=8.5cm,angle=0}
\figcaption
[f4.ps]{The central and radius=0.8\arcsec\
spectra for NGC~1399 (the two upper black lines) and the template
convolved with the best-fit velocity profiles (the two red lines).
The bottom spectrum is the template.
\label{fig4}}
\vskip 10pt

We then extract spectra at different spatial positions. The radii of
the extractions (in arcseconds) are 0.0, 0.10, 0.20, 0.30, 0.46,
0.81, and 1.62. Figure 5 plots the first and second moments as a function
of position along the slit. We have extracted the kinematics using
both the maximum penalized likelihood and using a Fourier
cross-correlation quotient technique (FCQ; Bender 1990). Both sets of
points are shown in Figure 5 and the results are similar. There are a
few positions where the differences are larger than statistical. For
example, at $+0.8$\arcsec, we find significant differences in the
dispersion measured between both techniques. This difference is
understandable given the double-peaked nature of the line-of-sight
velocity distribution (LOSVD) that we discuss in Section 3. When the
LOSVD is not unimodal, the way in which the velocity centroid is
measured can be very sensitive to fitting procedure. For those
positions where the velocity profile is unimodal, the two techniques
agree well.  In Fig.~ 5, we also include a symmetrized set of points
(the red line). Since the galaxy models that we use are axisymmetric,
in order to provide the highest S/N we symmetrize the fit
to the velocity profile. Details of this symmetrization are given in
Pinkney et al. (2003). We use the symmetrized values in the dynamical
models.

\vskip 10pt
\psfig{file=f5.ps,width=8.5cm,angle=0}
\figcaption
[f5.ps]{The first two moments of the velocity
profile as a function of position along the slit for NGC~1399. The
points with uncertainties are unsymmetrized. The filled black circles
come from the non-parametric estimate of the velocity profile, and the
open blue circles come from FCQ (Bender 1990). The red line is a
symmetrized version of the kinematic measurements which are
used in the dynamical modeling.
\label{fig5}}
\vskip 10pt 

One sees a dramatic increase in the dispersion to about 500~\kms\ at
0.5\arcsec\ on both sides of the galaxy. The dispersion then drops
toward the center to about 270~\kms, which is around the instrumental
sigma. However, by just using the dispersion alone, one does not get
the complete picture. At 0.5\arcsec, the velocity profile shows a
dramatic double peak. Figure 6 shows the velocity profiles as
determined by a symmetrized fit to opposite sides of the galaxy at the
labelled radii. As one moves to larger or smaller radii, the 
double-hump feature at $r=0.5$\arcsec\ slowly goes away until one gets a
nearly Gaussian profile. Furthermore, the double hump appears on both
sides of the center at $r=0.5$\arcsec.  Thus, the feature appears to be
robust. 

This feature and the drop in the dispersion toward the center suggest
an orbital distribution strongly dominated by tangential orbits. If
the orbital distribution consisted of stars only on circular orbits
(but with random orientations), then as one gets closer to the center,
the measured projected dispersion would drop to zero since all orbits
would have no radial component. Obviously, a disk is one mechanism
that would cause a drop in the central dispersion; but if the disk is
seen edge-on, there would be an obvious signature in the rotation
which is not seen. If the disk is face-on, it would be nearly
impossible for that configuration to cause the double-humped nature
seen in the LOSVDs. We are left to conclude that a stellar disk cannot
be the cause for the central dispersion drop. However, one can also
obtain a drop in the dispersion depending on the shape of the mass
density profile. But given the double-humped nature of the velocity
profile further out, it is likely that NGC~1399 is dominated by
tangential orbits near the center. In fact, the dynamical models
discussed below show the need for tangential orbits. Gebhardt et
al. (2003) discuss the signature of tangential orbits seen in other
galaxies. Thus, it appears that the largest galaxies, and hence those
with the flattest central density profiles, show the strongest amount
of tangential bias in the orbital distribution, with NGC~1399 being an
extreme example of this phenomenon.

Detailed inspections of the {\it HST} image and residual image (Lauer
et al. 2005) show nothing at $r=0.5$\arcsec.  As stated previously,
Houghton et al. (2006) find a flattened component in their adaptive
optics $K$-band image, with a similar radial extent. Clearly,
additional high-quality images would be worthwhile.

\subsection{Ground-Based Spectra}

There are ground-based spectra from two sources. Saglia et al. (2000)
published detailed models of NGC~1399 based on long-slit data and
include Gauss-Hermite polynomial expansion up to h4. Graham et
al. (1998) also provide long-slit data with measures of the first two
moments only. The comparison between the two datasets is excellent and
there is little difference in which one is used for the dynamical
models. However, Saglia et al. report additional information on the
shape of the velocity profile (4 moments compared to the 2 moments of
Graham et al.).  Since part of our goal is to constrain the stellar
orbital structure, it is better to use as much velocity profile shape
information as possible. Thus, for the following analysis, we use only
the Saglia et al. dataset.

\section{Dynamical Models}

The dynamical models that we use are based on orbit superpositions.
These are described in detail by Gebhardt et al. (2003), Thomas et
al. (2004), and Thomas et al. (2005). We will therefore not discuss
these models here, other than to provide our model parameters for
NGC~1399. Complete discussions of similar models are also given by
Cretton et al. (1999), Verolme et al. (2002), Valluri et al. (2004),
and Cappellari et al. (2006).

The models that we use for NGC~1399 have 20 radial and 5 angular
bins. Our orbit sampling has 20 energy bins, 40 angular momentum bins
(in the $z$ direction), and at least 15 bins for the third integral (see
Thomas et al. 2005 for a full description of the orbit sampling). We
only need to run models with one sign of angular momentum and then
double the number of orbits by flipping the individual velocity
profiles about zero velocity. The total number of orbits we have for
each orbit library is around 10000.  This orbit library is two times
higher than we generally use. However, we see no difference in the
results when using the smaller library.

The free parameters in the models are black hole mass, mass-to-light
ratio profile, and inclination. We use an edge-on projection for the
models shown below. Previous analysis (Gebhardt et al. 2003) shows that
different projections have little effect on the black hole
mass. Furthermore, since NGC~1399 is nearly round, one could even
consider spherical models (as in Houghton et al. 2006), which would
minimize projection effects. Figure 7 plots the $\chi^2$ versus black
hole mass, marginalized over the mass-to-light ratio. There is a
well-defined minimum and we exclude the zero black hole mass with a
$\Delta\chi^2=20$ (or $>99$\%). Since we marginalize over
mass-to-light ratio, our $1\sigma$ (68\%) confidence limit corresponds to
$\Delta\chi^2=1.0$. Thus, we find a best-fit black hole mass of
$(5.1 \pm 0.7) \times 10^8$~$\Msun$ (marginalized over $M/L$) 
and the best-fit $M/L_R$ of
$5.2\pm0.4$ (marginalized over black hole mass). In Figure 8 we plot
the two-dimensional $\chi^2$ contours for black hole mass and
mass-to-light ratio.

\vskip 10pt
\psfig{file=f6.ps,width=8.5cm,angle=0}
\figcaption
[f6.ps]{Line-of-sight velocity distributions
from STIS. These LOSVDs are the fit to both sides of the galaxy at the
specified radii; however, the LOSVD is flipped about the systemic
velocity, for the opposite side (i.e., this uses the axisymmetric
assumption). From 0.3\arcsec\ to 0.5\arcsec, the velocity profile becomes
double-peaked, which is the reason for the increase in the
measured second moment at this location in Fig. 5.
\label{fig6}}
\vskip 10pt

\vskip 10pt
\psfig{file=f7.ps,width=8.5cm,angle=-90}
\figcaption
[f7.ps]{$\chi^2$ versus black hole mass
marginalized over $M/L$ (left), and versus M/L marginalized over black
hole mass (right). The total number of parameters used in the fit is
131, but due to correlation between LOSVD bins, the effective number
is smaller.
\label{fig7}}
\vskip 10pt

\vskip 10pt
\psfig{file=f8.ps,width=8.5cm,angle=0}
\figcaption
[f8.ps]{Contours of $\chi^2$ as a function of
black hole mass and mass-to-light ratio. Each point represents a
particular model. The contours represent the 68, 90, 95, and 99\%
confidence for one degree-of-freedom, implying $\Delta\chi^2=1.0, 2.7,
4.0,$ and 6.6. The circled point is the model that has the minimum value.
\label{fig8}}
\vskip 10pt

Figure 9 plots the comparison between the first four Gauss-Hermite
coefficients of the data and the models. This plot can only be used
for a visual examination of how well the data are fitted, and a
statistical evaluation requires comparison with $\Delta\chi^2$, as inf
Figs 7 and 8.  Furthermore, the models are fitted using more
information than shown in Fig.~9 since we fit the full LOSVDs as
opposed to Gauss-Hermite coefficients. This fitting is especially
important for NGC~1399 since the LOSVDs are significantly
non-Gaussian.  In Figure 9 we plot three models: our best-fit model, a
model with no black hole, and a model with twice the best-fit mass (so
at $10^9~\Msun$). The $\chi^2$ difference of the two models compared to
the best fitted model is 10--12.

\vskip 10pt
\psfig{file=f9.ps,width=8.5cm,angle=0}
\figcaption
[f9.ps]{Comparison of data and models for the
first four Gauss-Hermite coefficients. The solid points represent the
STIS data, and the open points are ground-based measurements. The
lines are from three different models, with solid line for the STIS
data and the dashed line for ground-based. The black lines are from
the best-fit model, the green line is for the no black hole case, and
the blue line is for a mass that is twice the best-fit mass. The
dynamical models are fitted to the LOSVDs directly, so the comparison
with the Gauss-Hermite is only to provide a visual inspection of how
well we fit the data.
\label{fig9}}
\vskip 10pt

Using the integrated dispersion along the major axis out to an
effective radius (from the Saglia et al. 2000 data) provides a
dispersion of 337~\kms. The black hole mass in NGC~1399 is about a
factor of 2.5 below that expected from the BH/$\sigma$ correlation.

The strong tangential anisotropy seen in NGC~1399 is among the most
extreme seen in any galaxy to date. It is already clear in Fig. 5 that
tangential orbits dominate, but we also have a measure from the
dynamical models. Fig.~10 plots the radial to tangential dispersions
as a function of radius for all position angles in the galaxy. From
radii 0.1\arcsec--0.5\arcsec, the model becomes highly tangential,
with the ratio of the internal dispersions of the radial and
tangential components $\sigma_r/\sigma_t$ around 0.3. We can compare
the NGC~1399 orbital structure to those presented in Gebhardt et
al. (2003). There are many galaxies that have this amount of
tangential anisotropy in the central region, but none have such a
large radial extent. In fact, NGC~1399 is unique in that the central
bin is isotropic---which is rare in the Gebhardt et al. sample---but
then quickly becomes tangential outside the center.

There is important information from the position angles where we do
not have data. Even though there are no kinematic constraints there,
these offset axes have an effect in projection on the major-axis
kinematics, in particular near the center. Thus, there are indirect
kinematic constraints. Fig.~10 shows that the orbital structure along
these offset axes show a structure very similar to that along the
major axis.

\vskip 10pt
\psfig{file=f10.ps,width=8.5cm,angle=0}
\figcaption
[f10.ps]{Ratio of the radial to tangential second
moment of the velocity distribution for the best-fit BH model (top)
and the zero BH model (bottom). The solid line is along the major
axis, for which we have data. The dotted lines are along the other
four position angles in the model, for which we do not have data. We
have defined $\sigma_t = \sqrt{(\sigma_\theta^2+\sigma_\phi^2)/2}$,
where $\theta$ and $\phi$ are the standard spherical coordinates.  Our
best-fit model is the top panel. The solid red line is the ratio from
Houghton et al. (2006) for their best-fit model with a black hole of
$1.2 \times 10^9~\Msun$.
\label{fig10}}
\vskip 10pt

Houghton et al. (2006) find similar results for the orbital
structure.  Their ratio of radial to tangential dispersions is plotted
as the red line in Fig.~10. In both panels, we only plot the ratio for
their best-fit model, which has a black hole of $1.2\times10^9\Msun$.
While they do not find the extreme amount of tangential anisotropy
that we find for our best-fit model, the trend is very similar. Given
the better spatial information for the kinematics in our data, it is
not a surprise that we find a stronger change in the anisotropy.

\section{Uncertainties from the Orbit-Based Models}

For NGC~1399, we measure the black hole mass with 14\% accuracy.
Houghton et al. (2006) present dynamical models for NGC~1399 based on
kinematics obtained on the VLT with adaptive optics. Using
orbit-superposition models, they find a black hole mass of
$1.2(+0.5,-0.6)\times10^9~\Msun$, a 50\% accuracy. They also find
strong tangentially biased orbits in the central regions, which is
very similar to what we find (as plotted in Fig.~9). Statistically,
there is no concern since the two black hole masses are different by
only $1\sigma$. In fact, the Houghton et al. mass is consistent with
zero at $2\sigma$, so any black hole mass that we measure would be
consistent. The question, however, is why we provide an uncertainty
that is nearly ten times smaller than what they find. The answer is
most likely a combination of the data quality and differences in the
dynamical modeling, which we describe below. 

A similar comparison of the uncertainties can be made for other
galaxies with black hole mass estimates. The published uncertainties
range from 10\% to over 50\% (e.g., Tremaine et al. 2002), with most
of these based on orbit-based models from two groups (the Nuker and
the Leiden groups). However, Valluri et al. (2004) and Houghton et
al. (2006) both use orbit-based models and find substantially larger
uncertainties for the particular galaxies they model. Some of this is
due to the data that is being used but part of it is in the details of
the dynamical models. The most rigorous tests for recovery of the
black hole mass and uncertainties is in Siopis et al. (2007), where
they find that when using proper observational uncertainties, the
orbit-based models provide robust estimate of the black hole mass and
the uncertainties (also shown in Gebhardt 2003). As an example, one
can compare the black hole mass uncertainty for two of the best
measured stellar dynamical cases, our Galaxy and M32. Summarized in
Ghez et al. (2005) and Schodel et al. (2003), the black hole mass in
our Galaxy is known to 3--6\%, just over a factor of two better than
what we find in NGC~1399. Even though the Galaxy black hole is
significantly more spatially-resolved compared to NGC~1399, the
uncertainty is driven by the small number of stars with either radial
velocities or proper motions, whereas the signal-to-noise of the
central NGC~1399 is high enough that the uncertainty is driven mainly
by the spatial resolution. Thus, the relative accuracy of the black
hole masses is consistent. For M32, Verolme et al. (2002) measure the
black hole mass to 20\% accuracy for three degrees of freedom, and
about 10\% when using similar statistics as used for NGC~1399 (one
degree of freedom and marginalizing over the other parameters). Given
the relative distances, black hole masses and velocity dispersions,
the on-sky black hole sphere of influence in M32 is about $2\times$
smaller than in NGC~1399. Thus, the relative accuracies of the black
hole masses in this case is consistent as well.

However, there is a significant inconsistency with the uncertainty
measured here and in Houghton et al. (2006) for NGC~1399. For our
observations, the point spread function (PSF) of STIS is well
represented by an Airy function with most of the power in a single
Gaussian with FWHM=0.07\arcsec. Our central spectral element for
NGC~1399 is a $0.2\times0.1$\arcsec\ box. The PSF of the AO data from
Houghton et al. is complicated and they represent it as a double
Gaussian, with a strehl ratio of 30\% and
FHWM=0.15\arcsec. Furthermore, their PSF is simulated since the star
they use to provide the AO correction is 18\arcsec\ away from the
center of NGC~1399, but this is probably only adds a small additional
uncertainty on the PSF. Their slit is 0.17\arcsec\ wide. Convolving
both central spectral elements with the PSF shows that the STIS data
is about 50\% better than the AO data in terms of spatial resolution.
However, the main difference is due to the low strehl ratio of the AO
data. Since NGC~1399 has a relatively flat core, the 30\% strehl
causes light from larger radii to have a significant contribution to
the central spectral element. This effect is taken into account in
their modelling. Given the better PSF and strehl of STIS, the
uncertainties on the black hole are better by an appreciable
amount. The other main observational difference is the spectral
range. The STIS data uses the H+K and G-band regions, and Houghton et
al. use the CO-bandhead at 2.3~$\mu$m. Silge \& Gebhardt (2003) show
the complication that arise when using the bandhead and that the main
effect is to limit the accuracy of the LOSVD. Whether this effect is
part of the difference in the black hole accuracy is difficult to
ascertain, but could potentially be important.

The other important difference is the approach of the dynamical
models.  We both use orbit-based models, but we fit the LOSVD bins and
they fit basis functions as a representation of the LOSVD. The
advantage of the basis function is that they are mathematically
uncorrelated, and the LOSVD bins are correlated. This may have some
effect on the uncertainties as discussed in Magorrian (2006).  The
correlation of the LOSVD bins---and the similar correlation of
Gaussian-Hermite polynomial coefficients---would effect all black hole
mass uncertainties from stellar dynamics that have been
published. However, Gebhardt (2003) find that the uncertainties
estimated from the orbit-based models are accurate, based on bootstrap
simulations. Given the intrinsic scatter in black hole mass
correlations to host properties is close to zero, increasing the mass
uncertainties will push the intrinsic scatter to yet smaller
values. For NGC~1399, the differences in data quality appear to be
responsible for the difference in black hole mass
uncertainty. However, resolution of this difference in the modeling
approach will likely require a re-analysis of some of the data and
models.

\section{Discussion}

We have carefully examined the morphology around the radius where the
tangential orbits dominate, but we find no obvious feature.  There are
no changes in the surface brightness profile, the color profile, and
the ellipticity profile. A possible explanation for the tangential
orbits could have been a torus of material, as has been proposed to
explain hollow core galaxies (Lauer et al. 2002). A torus would also
manifest itself in the orbital structure in the offset axes. Since the
orbital structure appears to be similar along all position angles, we
argue that the tangential structure is independent of
angle. Furthermore, there is no net streaming motion measured in the
LOSVD, which argues that a disk is not the explanation.  More likely,
the cause could simply be a lack of radial orbits or an enhancement of
tangential orbits.

Using an integrated velocity dispersion of 337~\kms, the black hole
mass in NGC~1399 is a factor of 2.5 below that expected from the
BH/$\sigma$ correlation, and a factor of 2.0 below that expected from
the correlation with luminosity (Lauer et al. 2007). Saglia et
al. (2000), with much worse spatial resolution, find an upper limit on
the black hole mass that is consistent with our mass. It is possible
that the tangential orbits and the low black hole mass are related.
NGC~1399 does inhabit a special environment by being at the center of
the Fornax cluster. Whether more frequent accretion and mergers play a
role in shaping its black hole mass is unknown, and it would be
worthwhile to test whether binary black hole interactions could cause
both the tangential orbits and relatively low black hole
mass. However, the low black hole mass could reflect the intrinsic
scatter in the BH/$\sigma$ correlation, with NGC~1399 being near the
bottom edge of the observed scatter.

A possible scenario is to have a stellar cluster fall into NGC~1399 on
essentially a purely radial orbit. In this case the cluster hits the
black hole head-on and an equal number of stars pass to one side
and the other side, causing no net rotation. However, this would cause
there to be a preferred axis for the tangential orbits, and we see it
independent of angle. If, however, the stellar cluster is quite large
(i.e., around the size of the region of tangential anisotropy), then
the stars should distribute themselves in a spherical pattern. Those
stars that get near to the black hole --- the ones on radial orbits ---
tend to be ejected or accreted, leaving dominance of tangential
orbits. However, the extreme amount of tangential orbits in NGC~1399 
needs to be compared to a detailed simulation.

\acknowledgements

We are grateful for the hospitality provided by the Observatories of
the Carnegie Institute of Washington. KG is grateful to Andreas
Burkert for discussions on the nature of the tangential orbits, to
Jens Thomas for numerous discussion of the modelling code, and to John
Magorrian for many stimulating discussions. KG gratefully acknowledges
NSF CAREER grant AST 03-49095. This publication is based on
observations made with the NASA/ESA {\it Hubble Space Telescope},
which is operated by the Association of Universities for Research in
Astronomy, Inc., under NASA contract NAS 5-26555.  Financial support
was provided by NASA grants GO-5990 and GO-8214 from the Space
Telescope Science Institute.


\begin{references}

\reference{} Barth, A., Ho, L.C., Sargent, W. 2002, \apjl, 566, L13
\reference{} Bender, R. 1990, A\&A, 229, 441
\reference{} Cappellari, M., et al. 2006, \mnras, 366, 1126
\reference{} Cretton, N., de Zeeuw, P. T., van~der~Marel, R. P., \&
Rix, H.-W. 1999, \apjs, 124, 383
\reference{} Dressler, A. 1984, \apj, 281, 512
\reference{} Dressler, A. 1989, IAU Symposium, 134, 217
\reference{} Fabian, A. 1999, \mnras, 398, L6
\reference{} Ferrarese, L., \& Merritt, D. 2000, \apjl, 539, L9
\reference{} Gebhardt, K., et al. 1996, \aj, 112, 105
\reference{} Gebhardt, K., et al. 2000, \apj, 539, L13
\reference{} Gebhardt, K., et al. 2003, \apj, 583, 92
\reference{} Gebhardt, K. 2003, in ``Carnegie Observatories Astrophysics 
Series, Vol. 1: Coevolution of Black Holes and Galaxies, ed. L.C. Ho 
(Pasadena: Carnegie Observatories)
\reference{} Ghez, A. et al. 200
\reference{} Graham, A., Colless, M., Busarello, G., Zaggia, S., \& Longo, G.
1998, A\&AS, 133, 325
\reference{} Houghton, R., Magorrian, J., Sarzi, M., Thatte, N., Davies, R.,
Krajnovic, D. 2006, \mnras, 367, 2
\reference{} Kormendy, J. 1993, in The Nearest Active Galaxies, ed. J. Beckman, 
L. Colina, \& H. Netzer (Madrid:CSIC), 197
\reference{} Kormendy, J., \& Richstone, D. 1995, \araa, 33, 581
\reference{} Lauer, T., et al. 2002, \aj, 124, 1975
\reference{} Lauer, T., et al. 2005, \aj, 129, 2138
\reference{} Lauer, T., et al. 2007, \apj, 662, 808
\reference{} Leitherer et al. 1996, PASP, 108, 996
\reference{} Magorrian, J.,~\etal\ 1998, \aj, 115, 2285
\reference{} Magorrian, J. 2006, \mnras, 373, 425
\reference{} Pinkney, J. et al. 2003, \apj, 596, 903
\reference{} Robertson, B., Hernquist, L., Cox, T., Di Matteo, T., Hopkins, P., 
Martini, P., \& Springel, V. 2006, \apj, 641, 90
\reference{} Saglia, R.P., Kronawitter, A., Gerhard, O., \& 
  Bender, R. 2000, \aj, 119, 153
\reference{} Schodel, R, Ott, T., Genzel, R., Eckart, A., Mouawad, N.,
\& Alexander, T. 2003, \apj, 596, 1015
\reference{} Silge, J. \& Gebhardt, K. 2003, \aj, 125, 2809
\reference{} Silk, J., \& Rees, M. J. 1998 A\&A, 331, L1
\reference{} Siopis, C. et al. 2007, submitted.
\reference{} Springel, V., Di Matteo, T., \& Hernquist, L. 2005, \mnras, 361, 776
\reference{} Thomas, J., Saglia, R., Bender, R., Thomas, D., Gebhardt, K., 
Magorrian, J., \& Richstone, D. 2004, \mnras, 353, 391
\reference{} Thomas, J., Saglia, R., Bender, R., Thomas, D., Gebhardt, K., 
Magorrian, J., Corsini, E., \& Wegner, G. 2005, \mnras, 360, 1355
\reference{} Tonry, J. L., Dressler, A., Blakeslee, J. P., Ajhar,
E. A., Fletcher, A. B., Luppino, G. A., Metzger, M. R., \& Moore,
C. B. 2001, \apj, 546, 681
\reference{} Tremaine, S.,~\etal\ 2002, \apj, 574, 740
\reference{} Valluri, M., Merritt, D., \& Emsellem, E. 2004, \apj, 602, 66
\reference{} Verolme, E.,~\etal\ 2002, \mnras, 335, 517
\reference{} Woodgate, B. et al. 1998, \pasp, 110, 1183

\end{references}
\end{document}